\documentstyle[12pt]{article}
\begin{document}





\font\trm = cmr10 scaled \magstep3
\font\srm = cmr10 scaled \magstep2
\font\urm = cmr10 scaled \magstep2
\font\bbf = cmssbx10 
\font\bbb = cmbx12
\font\nbf = cmbx10
\font\gbf = cmtt10 scaled \magstep4 \font\tbf=cmtt10 scaled \magstep2
\font\sevenrm = cmr8
\scriptscriptfont0 =\scriptfont0
\scriptscriptfont1 =\scriptfont1
\font\grb = cmmi10 scaled \magstep3

\def\lin{\bigskip\centerline{\vrule width 5cm height0.4pt}\bigskip}
\def\d{\partial}
\def\dh{\mathop{\vphantom{\odot}\hbox{$\partial$}}}
\def\dl{\dh^\leftrightarrow}
\def\sqr#1#2{{\vcenter{\vbox{\hrule height.#2pt\hbox{\vrule width.#2pt 
height#1pt \kern#1pt \vrule width.#2pt}\hrule height.#2pt}}}}
\def\w{\mathchoice\sqr45\sqr45\sqr{2.1}3\sqr{1.5}3\,} 
\def\fii{\varphi}
\def\eps{\varepsilon}
\def\hq{\hbar}
\def\lo{\cal L_+^\uparrow}
\def\psq{{\overline{\psi}}}
\def\pp{\psi ^\prime}
\def\ppq{\overline{\psi ^\prime}}
\def\sp{\vec\sigma\cdot\vec p}
\def\pdh{(2\pi )^{-3/2}}
\def\ps{\hbox{\rm\rlap/p}}
\def\ec{{e\over c}}
\def\=d{\,{\buildrel\rm def\over =}\,}
\def\iix{\int d^3x\,}
\def\iip{\int d^3p}
\def\inx{\int d^3x_1\ldots d^3x_n}
\def\H{{\cal H}}
\def\F{{\cal F}}
\def\N{\hbox{\bbf N}}
\def\A{\hbox{\bbf A}}
\def\xn{\vec x_1,\ldots ,\vec x_n\,}
\def\vxp{\vec x\,'}
\def\V{\hbox{\bbf V}}
\def\S{\hbox{\bbf S}}
\def\U{{\hbox{\bbf U}}}
\def\HH{\hbox{\bbf H}}
\def\Q{\hbox{\bbf Q}}
\def\i3p{\p32\int d^3p}
\def\psm{\psi ^{(-)}}
\def\psp{\psi ^{(+)}}
\def\px{\vec p\cdot\vec x}
\def\pqp{\overline{\psi}^{(+)}}
\def\pqm{\overline{\psi}^{(-)}}
\def\vq{\overline v}
\def\uq{\overline u}
\def\iep{\int {d^3p\over 2E}}
\def\ipd{\int d^4p\,\delta (p^2-m^2)}
\def\ds{\hbox{\rlap/$\partial$}}
\def\pe{\sqrt{\vec p^2+m^2}}
\def\dsx{\hbox{\rlap/$\partial$}_x}
\def\itn{\int\limits_{-\infty}^{+\infty}dt_1\int\limits_{-\infty}^{t_1}
dt_2\cdots\int\limits_{-\infty}^{t_{n-1}}dt_n}
\def\ipn{\int d^3p_1\cdots\int d^3p_{n-1}\, }
\def\As{A\hbox to 1pt{\hss /}}
\def\np4{\int d^4p_1\cdots d^4p_{n-1}\, }
\def\Sr{S^{\rm ret}}
\def\gp{\vec\gamma\cdot\vec p}
\def\te{\vartheta}
\def\tr{{\rm tr}\, }
\def\Sa{S^{\rm av}}
\def\qs{\hbox{\rlap/q}}
\def\supp{{\rm supp}\, }
\def\Tr{{\rm Tr}\, }
\def\Im{{\rm Im}\, }
\def\sgn{{\rm sgn}\, }
\def\cau{{1\over 2\pi i}}
\def\P{\,{\rm P}}
\def\Re{{\rm Re}\, }
\def\iinf{\int\limits_{-\infty}^{+\infty}}
\def\kx{\vec k\cdot\vec x}
\def\io{\int{d^3k\over\sqrt{2\omega}}\,}
\def\nx4{\int d^4x_1\ldots d^4x_n\, }
\def\xnn{x_1,\ldots ,x_n}
\def\xnm{x_1,\ldots ,x_{n-1},x_n}
\def\Dr{D^{\rm ret}}
\def\Da{D^{\rm av}}
\def\kon#1#2{\vbox{\halign{##&&##\cr
\lower4pt\hbox{$\scriptscriptstyle\vert$}\hrulefill &
\hrulefill\lower4pt\hbox{$\scriptscriptstyle\vert$}\cr $#1$&
$#2$\cr}}}
\def\lra{\longleftrightarrow}
\def\konv#1#2#3{\hbox{\vrule height12pt depth-1pt}
\vbox{\hrule height12pt width#1cm depth-11.6pt}
\hbox{\vrule height6.5pt depth-0.5pt}
\vbox{\hrule height11pt width#2cm depth-10.6pt\kern5pt
      \hrule height6.5pt width#2cm depth-6.1pt}
\hbox{\vrule height12pt depth-1pt}
\vbox{\hrule height6.5pt width#3cm depth-6.1pt}
\hbox{\vrule height6.5pt depth-0.5pt}}
\def\konu#1#2#3{\hbox{\vrule height12pt depth-1pt}
\vbox{\hrule height1pt width#1cm depth-0.6pt}
\hbox{\vrule height12pt depth-6.5pt}
\vbox{\hrule height6pt width#2cm depth-5.6pt\kern5pt
      \hrule height1pt width#2cm depth-0.6pt}
\hbox{\vrule height12pt depth-6.5pt}
\vbox{\hrule height1pt width#3cm depth-0.6pt}
\hbox{\vrule height12pt depth-1pt}}
\def\es{\hbox{\rlap/$\varepsilon$}}
\def\ks{\hbox{\rlap/k}}
\def\konw#1#2#3{\hbox{\vrule height12pt depth-1pt}
\vbox{\hrule height12pt width#1cm depth-11.6pt}
\hbox{\vrule height6.5pt depth-0.5pt}
\vbox{\hrule height12pt width#2cm depth-11.6pt \kern5pt
      \hrule height6.5pt width#2cm depth-6.1pt}
\hbox{\vrule height6.5pt depth-0.5pt}
\vbox{\hrule height12pt width#3cm depth-11.6pt}
\hbox{\vrule height12pt depth-1pt}}
\def\grad{{\rm grad}\, }
\def\diw{{\rm div}\, }
\def\eh{{\scriptstyle{1\over 2}}}
\def\intt{\int dt_1\ldots dt_n\,}
\def\tnn{t_1,\ldots ,t_n}
\def\dett{{\rm det}\,}
\def\lap{\bigtriangleup\,}
\def\HHe{\hbox{\bex H}}
\font\bex=cmbx7
\def\Wiout{W_{{\rm\scriptstyle in}\atop{\rm\scriptstyle out}}}
\def\Win{W_{\rm in}}
\def\Wout{W_{\rm out}}
\def\poi{\cal P _+^\uparrow}
\def\gu{\underline{g}} 
\def\hu{\underline{h}}
\def\i{{\rm int}}
\def\su{\sum_{n=1}^\infty}
\def\suu{\sum_{n=0}^\infty}
\def\c{{\rm cl}}
\def\e{{\rm ext}}
\def\snn{\su {1\over n!}\nx4}
\def\r{{\rm ret}}
\def\a{{\rm av}}
\def\ra{{{\rm\scriptstyle ret}\atop{\rm\scriptstyle av}}}
\def\dsu{\mathop{\bigoplus}}            
\def\m3{{\mu_1\mu_2\mu_3}}
\def\itt{\int\limits_{t_1}^{t_2}}
\def\gs{\,{\scriptstyle{>\atop <}}\,}
\def\ga{\,{\scriptstyle{>\atop\sim}\,}}
\def\Seh{{\cal S}^\eh_\eh}
\def\SS{{\cal S}({\bf R^4})}
\def\co{{\rm Com}}
\def\ml{{m_1\ldots m_l}}
\def\ik{{i_1\ldots i_k}}
\def\jt{{j_1\ldots j_t}}
\def\is{{i_{k+1}\ldots i_s}}
\def\js{{j_{t+1}\ldots j_s}}
\def\xnr{x_1-x_n,\ldots x_{n-1}-x_n}
\def\qm{q_1,\ldots q_{m-1}}
\def\pn{p_1,\ldots p_{n-1}}
\def\xll{x_l-x_{l+1}}
\def\xmm{x_1,\ldots x_{n-1}}
\def\ph{{\rm phys}}
\def\nab{\bigtriangledown}
\def\p{{(+)}}
\def\ul{\underline}
\def\tu{\tilde u}


\textwidth=14.5cm
\textheight=20cm
\pagestyle{plain} 
\parindent0cm

\newcommand{\ttbs}{\char'134}
\newcommand{\AmS}{{\protect\the\textfont2
  A\kern-.1667em\lower.5ex\hbox{M}\kern-.125emS}}
\def\beq{\begin{equation}}
\def\eeq{\end{equation}}
\def\bea{\begin{eqnarray}}
\def\eea{\end{eqnarray}}
\language 1
\hyphenation{non-abel-ian Higgs equi-va-lent in-ter-est-ing re-de-fi-ni-tion}


\thispagestyle{empty}
\setcounter{footnote}{0}
{\small 
{\small\it  ITP-UH-4/97}\\
{\small\it ITP-SB-97-13}\\
\vbox to 1,5cm{ }
\centerline{\large\bf Polynomial Form of the Stueckelberg Model $^*$}
\vskip 2.0cm
\centerline{\bf Norbert Dragon,}
\vskip 0.2cm
\centerline{\it
Institut f\"ur Theoretische Physik,} 
\centerline{\it Universit\"at Hannover,}
\centerline{\it Appelstrasse 2, D-30167 Hannover}
\vskip 0,2cm
\centerline{\bf Tobias Hurth and Peter van Nieuwenhuizen} 
\vskip 0.2cm
\centerline{\it
Institute for Theoretical Physics,} 
\centerline{\it State University of New York at Stony Brook }
\centerline{\it Stony Brook, New York 11794-3840}
\vskip 1,0cm
{\bf Abstract.} - 
The Stueckelberg model for massive vector fields is cast into a BRS invariant, polynomial
form. Its symmetry algebra simplifies to an abelian gauge symmetry which is sufficient to
decouple the negative norm states. The propagators fall off like $1/k^2$ and
the Lagrangean is polynomial but it is not powercounting renormalizable due to derivative
couplings.

\vskip 3,5cm
$^*)$To appear in the Proceedings of the 30th International Symposium
Ahrenshoop on the Theory of Elementary Particles, Nuclear Physics B,
       Proceedings Supplement, edited by D. L\"ust, 
       H.-J. Otto and G. Weigt.\\
       Research supported by NSF grant no 9309888 and the 
       Swiss National Science Foundation.

\newpage

The Stueckelberg model describes massive vector particles with gauge interactions. It has repeatedly
attracted attention \cite{Delbougho} as an alternative to the Higgs model and is particularly
interesting as long as the Higgs particle is not confirmed experimentally.

The model contains a set $A_\mu{}^a, \varphi^a$ of real vector and scalar fields ($a=1,\dots,{\rm dim}(G)$) 
with a gauge invariant kinetic energy and with the interactions  of a nonabelian gauge group $G$, which we
take to be simple. 
The coupling constant $g$ appears as a normalization $\frac{1}{g^2}$ in front of the action.
\beq
{\cal L}_{kin}=-\frac{1}{4g^2}F_{\mu\nu}{}^aF^{\mu\nu}{}^a
\eeq
\beq
F_{\mu\nu}{}^a=\partial_\mu A_\nu{}^a - \partial_\nu A_\mu{}^a -
f_{bc}{}^a A_\mu{}^b A_\nu{}^c
\eeq
The mass term for the vector fields is introduced in a gauge invariant way.
\beq
{\cal L}_{mass}=- \frac{m^2}{g^2} \;{\rm tr}\, ( A_\mu-U^{-1}\partial_\mu U)^2
\eeq
where $U=U(\varphi)$ is an abbreviation for the series
\beq
U=e^{\frac{1}{m}\varphi^a T_a}
\eeq
and $A_\mu$ is the matrix  
\beq
A_\mu = A_\mu{}^aT_a.
\eeq
The matrices $T_a$ are a normalized basis of a matrix representation of the Lie algebra of $G$.
\beq
[T_a,T_b]=f_{ab}{}^c T_c \qquad {\rm tr}\, T_a T_b = - \frac{1}{2}\delta_{ab}
\eeq
One easily confirms the invariance of ${\cal L}_{kin}$ and of ${\cal L}_{mass}$ under the
gauge transformation generated by 
\beq
V=e^{\lambda^a (x) T_a}.
\eeq
\bea
U \rightarrow U^\prime &=& U\cdot V\\
\nonumber
A_\mu \rightarrow A^\prime_\mu &=& V^{-1}\cdot(A_\mu+\partial_\mu)\cdot V
\eea
In particular one can choose the gauge 
\beq
\label{u-1}
V=U^{-1}.
\eeq
Thereby one gets rid of the scalar fields $\varphi^a$ and fixes the gauge symmetry and one may well
wonder whether  $\varphi^a$  and the gauge symmetry with arbitrary functions $\lambda ^a$
are trivial. In the gauge (\ref{u-1}) the Lagrangean for $A^\prime_\mu$
\beq
{\cal L}_{inv}={\cal L}_{kin}+{\cal L}_{mass}
\eeq
is the Lagrangean ${\cal L}_{YM, mass}$ of massive Yang Mills fields with a propagator
\bea
\nonumber
\label{massiv}
<\!{\rm T}A^\prime_\mu (x) A^\prime_\nu (0)\!>\!\!\!\!&\!\!=&\!\!\!\!\! -{\rm i}g^2\!\!\int\!\!\frac{d^4 k}{(2\pi)^4}e^{{\rm 
i}kx}
\frac{\eta_{\mu\nu}-\frac{k_\mu k_\nu}{m^2}}{k^2-m^2+{\rm i}\epsilon}\\
&&
\eea
which does not fall off like $1/k^2$. 

A different choice of the gauge, however, leads to well
behaved propagators where   \hbox{$<AA>$} and $<\varphi\varphi>$ fall off like  $1/k^2$. The additional
propagating degrees of freedom in $A$ and $\varphi$, which yield these improved propagators,
have to be compensated by anticommuting Fadeev-Popov ghosts $b^a$ and $c^a$. They decouple
from the three physical spin-1 modes in the vector field because ${\cal L}_{inv}$ and the
gauge fixing and ghost Lagrangean are invariant under BRS transformations $s$ \cite{BRS}.
Explicitly $s$ is given by
\bea
\nonumber
s\,A_\mu{}^a &=& \partial_\mu c^a - c^b A_\mu{}^cf_{bc}{}^a\\
\label{solve}
s\,\left ( e^{\frac{1}{m}\varphi^a T_a}\right )&=&\left (e^{\frac{1}{m}\varphi^a T_a}\right ) c^bT_b\\
\nonumber
s\, c^a &=& -\,c^bc^cf_{bc}{}^a\\
\nonumber
s\, b^a &=& B^a \\
\nonumber
s\, B^a &=& 0\ .
\eea
$B^a$ is an auxiliary field introduced to make the BRS transformation $s$  offshell nilpotent.
The transformation of $\varphi$ which follows from (\ref{solve}) is nonpolynomial
\beq
s\,\varphi^a = m\,c^a+ \frac{1}{2}c^c\varphi^b f_{bc}{}^a + \dots
\eeq
and also the Lagrangean is an infinite series because of terms such as
$$
{\cal L}_{inv}= \dots
-m^2{\rm tr\, } U^{-1}\partial_\mu U U^{-1}\partial^\mu U.
$$

The transformation of the fields and the Lagrangean can be simplified considerably.
It is well known \cite{CWZ}
that if one is given a set of fields which transform under a compact group $G$
one can redefine the fields\footnote{
The perturbatively evaluated $S$-matrix is unchanged by this field redefinition,
global transformation properties like existence of fixed points in the space of fields are
normally not preserved.} such that they consist of multiplets of tensor fields transforming 
linearly under a subgroup $H$ and Goldstone fields which transform like the coset $G/H$ under
group multiplication.
In the case at hand the scalar fields transform as Goldstone fields of $G$, i.e.
 $H=\mbox{1\hspace{-.3em}I}$,
and the vector fields $A_\mu$, which transform under the adjoint transformation
of $G$, can be replaced by $G$-invariant vector fields $\hat{A}_\mu$.
\bea
\nonumber
\hat{A}_\mu &=& U\cdot (A_\mu -U^{-1} \partial_\mu U )\cdot U^{-1}\\
&=& A_\mu -\frac{1}{m}\partial_\mu \varphi^a T_a + O(A_\mu\varphi, \partial_\mu \varphi \varphi)
\eea
The Lagrangean ${\cal L}_{inv}(A,\partial A,\varphi, \partial \varphi)$ actually depends only on $\hat{A}$
and $\partial \hat{A}$ and only in a polynomial form
\bea
\nonumber
{\cal L}_{inv}(A,\partial A,\varphi, \partial \varphi)&=&
{\cal L}_{inv}(\hat{A},\partial \hat{A},0 , 0)\\
&=&{\cal L}_{YM, mass}(\hat{A},\partial \hat{A}).
\eea
The ghost system is drastically simplified by a redefinition of the
ghost fields $c^a$ $\grave{\rm a}$ la Brandt \cite{brandt}.
\beq
\hat{c}^a = \frac{1}{m}s \varphi^a
\eeq
This is also an invertible field transformation which does not change the
perturbatively evaluated $S$-matrix. It casts the transformation of the scalar fields into an abelian
form with no remnants of the structure constants $f_{ab}{}^c$ of the nonabelian group $G$.
In terms of the fields $\hat{A}_\mu$ and $\hat{c}^a$ the BRS transformations are linear and read
\bea
\nonumber
s \hat{A}_\mu&=0 \phantom{A^a \quad s A}&\phantom{=0}\\
\label{abelian}
s \varphi^a &=m \hat{c}^a \quad s \hat{c}^a &=0\\
\nonumber
s b^a &=B^a \quad sB^a&=0 .
\eea
The BRS multiplets consist of pairs $(\varphi^a, \hat{c}^a)$, $(b^a, B^a)$ and singlets
$\hat{A}^a_\mu$. Therefore all BRS invariant local actions, i.e. all Lagrangeans which are invariant
up to derivatives, follow from the basic lemma (see e.g. \cite{dragon}). 
\beq
s {\cal L}= 0 \Leftrightarrow {\cal L}= {\cal L}_{phys}(\hat{A},\partial \hat{A})+ s (b^a X_a),
\eeq
where $X_a$ are real bosonic functions of all the fields and their derivatives. They have to have ghost 
number 0
and can otherwise be conveniently chosen because
the piece $s (b^a X_a)$ has no measurable effect on scattering amplitudes of physical states,
as long as a change of parameters in $X_a$ perturb these scattering amplitudes smoothly.

If, for example, we choose 
$X_a=X_a(\varphi,B)=\frac{1}{2}B^a-(\Box + m^2) \frac{\varphi^a}{m}$ then 
$\varphi^a$ are dipole fields \cite{Ferrara}. They contain two species of creation operators for spin-0 particles.
 One of the
creation operators, $a^\dagger$,  appears with the usual plane wave $e^{{\rm i}kx}$, it is not invariant under BRS 
transformation
but transformed into the creation operator $c^\dagger$ contained in $\hat{c}^a$, because $s\varphi=m\,\hat{c}$. Hence 
$sa^\dagger=m\,c^\dagger$.
 The other creation operator in $\varphi^a$, $d^\dagger$, 
appears with $x^0e^{{\rm i}kx}$. 
Since there is no $x^0e^{{\rm i}kx}$ term in $\hat{c}$, $s\varphi=m\,\hat{c}$ implies that $sd^\dagger=0$. This agrees with $sb=B=(\Box + m^2) \frac{\varphi}{m}$  because 
in $ (\Box + m^2) \varphi$ only a term $2ik^0d^\dagger e^{{\rm i}kx}$
is left, while $b$ contains a term $b^\dagger e^{{\rm i}kx}$ so that $sb^\dagger\propto d^\dagger$, and thus indeed $sd^\dagger=0$. The doublets $(b^\dagger,d^\dagger)$ and $(a^\dagger,c^\dagger)$ have opposite ghost numbers, and form a 
Kugo-Ojima quartet \cite{kugo}. 

These quartets of creation operators in the fields $\varphi, \hat{c}, b$ decouple from the space of physical states
which is generated by the vector field  $\hat{A}_\mu$. $\hat{A}_\mu$ has the
undesirable propagator (\ref{massiv}) if ${\cal L}_{phys}={\cal L}_{YM, mass}$ and 
if the functions $X_a$ are chosen to be independent of $\hat{A}_\mu$. 

One is tempted to add to ${\cal L}_{YM, mass}$ a piece $\frac{\lambda}{2g^2}(\partial_\mu A^\mu)^2$ 
to obtain a propagator  $<\!\hat{A}\hat{A}\!>$ which falls off like 
$\frac{1}{k^2}$ for $|k^2| \gg \frac{m^2}{\lambda}$. But this is achieved only at the expense
of a coupled negative norm state at mass $\frac{m}{\sqrt{\lambda}}$ \cite{Boer}. The invariance of the
action under the BRS transformations (\ref{abelian}) does not guarantee the decoupling of
this negative norm state because the term $\frac{\lambda}{2g^2}(\partial_\mu A^\mu)^2$
is not introduced as part of the gauge fixing and ghost sector $s(b^aX_a)$ of the Lagrangean.

One can, however, obtain the desired propagator for the massive 
vector field if one chooses
\beq
X_a = \frac{1}{g^2}(\frac{1}{2}B^a - \partial \hat{A}^a - (\Box + m^2)\frac{\varphi^a}{m}\,).
\eeq
This yields the following gauge fixing and ghost part
\bea
\nonumber
s(b^aX_a) &=&  \frac{1}{2g^2}(B^a - \partial \hat{A}^a - (\Box + m^2)\frac{\varphi^a}{m}\,)^2-\\ 
\nonumber
&&-\frac{1}{2g^2}(\partial \hat{A}^a + (\Box + m^2)\frac{\varphi^a}{m}\,)^2 +\\
&&
+\frac{1}{g^2} b^a(\Box + m^2)\hat{c}^a.
\eea
The field $B^a$ is auxiliary with the algebraic equation of motion
\beq
B^a = \partial \hat{A}^a + (\Box + m^2)\frac{\varphi^a}{m} \ .
\eeq
The ghosts $b^a$ and $\hat{c}^a$ are free! The $\hat{A}-\varphi$ sector does not contain
higher derivatives or dipoles because
\bea
\partial \hat{A}+ (\Box + m^2)\frac{\varphi}{m}&=&\partial(\hat{A}+\frac{1}{m}\partial \varphi) + m \varphi\\
\nonumber
&=&\partial \bar{A}+m \varphi
\eea
where
\beq
\bar{A}=\hat{A}+\frac{1}{m}\partial \varphi\ .
\eeq
If one writes the Lagrangean in terms of $\bar{A}$ and $\varphi$ one obtains finally
\bea
\nonumber
{\cal L}_{YM, mass}(\hat{A})+s(b^aX_a)=-\frac{1}{4g^2}F_{\mu\nu}{}^2 +&&\\
\nonumber
+\frac{m^2}{2g^2}(\bar{A}-\frac{1}{m}\partial \varphi )^2 -\frac{1}{2g^2}(\partial\bar{A}+m\varphi)^2+&&\\
+ {\cal L}_{ghost}+{\cal L}_{auxiliary} = {\cal L}_{(2)}+{\cal L}_{int}\ .&&
\eea
In particular the free Lagrangean ${\cal L}_{(2)}$ does not mix the fields ${\bar{A}}$ and
$\varphi$. Dropping complete derivatives it takes the form
\bea
\nonumber
g^2{\cal L}_{(2)}\!\!\!\!\!&=&\!\!\!\!\! \frac{1}{2}\bar{A}_\mu^a(\Box+m^2)\bar{A}^{\mu\,a} 
-\frac{1}{2}\varphi^a(\Box + m^2)\varphi^a +\\
+ b^a(\Box \!\!\!\!\! & +&\!\!\!\!\! m^2)\hat{c}^a + \frac{1}{2}(B^a-\partial \bar{A}^a-m\varphi^a)^2\ .
\eea
All propagators fall off like ${1}/k^2$. 

The states which are
generated by $\partial \bar{A}$ decouple because they are not BRS invariant and the
states generated by $\varphi +\frac{1}{m}\partial \bar{A}=s \frac{1}{m}b$ decouple because 
they are BRS trivial. Together with the states generated by $b^a$ and $\hat{c}^a$ they form BRS quartets \cite{kugo}
which decouple from the Hilbert space of physical states.

The interaction is contained in the interaction part of ${\cal L}_{YM, mass}$ where, however,
the field $\hat{A}$ is replaced by $\bar{A}-\frac{1}{m}\partial\varphi$.
\beq
{\cal L}_{int}= -\frac{1}{4g^2}({F_{\mu\nu}{}^a}(\bar{A}-\frac{1}{m}\partial\varphi))^2_{\,|_{int}}
\eeq
It contains derivative couplings of $\varphi$ up to dimension 8. 

The nonrenormalizability from the infinite
series $U(\varphi)$ has been tamed to a finite number of nonrenormalizable interactions
and the Stueckelberg model has been simplified considerably.

In fact, it has been simplified so much
that a sobering analysis of the physical amplitudes is possible. Physical states are generated by
the spin-1 part of the field $\hat{A}=\bar{A}-\frac{1}{m}\partial \varphi$. This field has only interactions
with itself, in fact it has just the Yang Mills interactions. The propagator of $\hat{A}$ follows from
the propagators of $A$ and $\varphi$
\beq
<\!\hat{A}\hat{A}\!> = <\!A A\! > + \frac{1}{m^2}<\!\partial \varphi \partial \varphi\!>
\eeq
and turns out to be just the one in (\ref{massiv}) we started from.
\bea
\nonumber
<\!{\rm T}\hat{A}_\mu (x) \hat{A}_\nu (0)\!>\!\!\!\!\!&=&\!\!\!\!\!\! -{\rm i}g^2\!\!\int\! \frac{d^4 k}{(2\pi)^4}e^{{\rm 
i}kx}
\frac{\eta_{\mu\nu}-\frac{k_\mu k_\nu}{m^2}}{k^2-m^2+{\rm i}\epsilon}\\
&&
\eea
In the polynomial formulation of the Stueckelberg model the propagator with bad
high energy behaviour is exchanged for well behaved propagators and derivative couplings. 
While such a reformulation is mathematically completely equivalent it may nevertheless
be more fruitful to think of the diagrams of massive Yang Mills theory in terms of
exchanged vectors and scalars. For example, when Veltman 
analyzed massive Yang Mills theories \cite{Veltmana} he could show the absence
of on-shell one loop divergencies in 1PI diagrams with more than 4 external spin-1 particles. He achieved his 
result by adding and mixing  free scalars in a 
very similar manner to the approach followed here.
However, he only claimed that the one-loop theory was 
power-counting renormalizable, not that it was multiplicatively renormalizable
\cite{Veltmanb}. At the two-loop level, he showed that replacing $k^2$ by $k^2+m^2$ in YM theory, keeping 
the rest of the Feynman rules as in the massless case, leads to non-unitarity. He also studied which contact 
terms are needed for unitarity (noting that $(\eta_{\mu\nu}-\partial_\mu \partial_\nu/m^2) (\Theta \ \Delta^{+/-})$ 
differs from the required $\Theta (\eta_{\mu\nu}-\partial_\mu \partial_\nu/m^2) \Delta^{+/-}$ by contact terms 
$\delta_{\mu 0} \delta_{\nu 0} \delta^3(x)$). However, he reached no definite conclusions at the two-loop 
level 
\cite{Veltmanc}. We stress that renormalization of off-shell Greens functions 
can be weakened to renormalization of S-matrix elements. It would be 
a formidable but interesting problem to analyze whether massive YM theory has 
a two-loop renormalizable S-matrix.
 
It is interesting to know whether by field rede\-finitions one can cast a model,
like the polynomial Stueckelberg model,
into a power counting renormalizable form. In the massive, abelian case this 
is possible, because
the field strengths $F_{\mu\nu}$ calculated from $\bar{A}$ and from 
$\bar{A}-\frac{1}{m}\partial \varphi$ coincide and the model is actually free.  If one introduces
additional matter, like a gauge invariant Dirac fermion $\Psi$, and adds a gauge invariant
interaction
\beq
{\cal L}_{matter}= \bar{\Psi}({\rm i}\gamma\partial- M)\Psi + 
\bar{\Psi}\gamma(\bar{A}-\frac{1}{m}\partial\varphi)\Psi
\eeq
then the apparent nonrenormalizable derivative coupling  $\bar{\Psi}\gamma\partial\varphi\Psi$
can be absorbed by a field redefinition
\beq
\psi=e^{\frac{\rm i}{m}\varphi }\Psi,
\eeq
\beq
{\cal L}_{matter}= \bar{\psi}(\gamma({\rm i}\partial + \bar{A}) -M)\psi.
\eeq
In the massive, nonabelian case this question was first analyzed in \cite{Cornwall}. It was shown that the 
criterion of tree-unitarity uniquely leads to the
Higgs model. (See also \cite{Hu} where it was shown using asymptotic BRS-invariance that the nonabelian 
Stueckelberg model fails to be power-counting renormalizable and that no field redefinitions exist to establish 
this property.) However, it would be nice  to determine by algebraic methods only whether for the self 
interacting nonabelian theory described by
the Stueckelberg model a renormalizable formulation is possible only if one introduces
additional scalar fields and additional couplings of the scalar fields such that the Stueckelberg
model becomes a subsector of the nonabelian Higgs model.

\end{document}